\documentclass[pre,twocolumn,aps,showkeys]{revtex4}

\usepackage{graphicx}
\usepackage{dcolumn}
\usepackage{bm}

\begin{document}


\title{Random close packing fractions of
lognormal distributions of hard spheres}

\author{Robert S. Farr}
 \affiliation{Unilever R\&D, Colworth Science Park, Sharnbrook, Bedford, UK, MK44 1LQ}
 \affiliation{The London Institute for Mathematical Sciences, 35a South St., Mayfair, London, UK.}
 \email{robert.farr@unilever.com}

\date{\today}

\begin{abstract}
We apply a recent one-dimensional algorithm for predicting 
random close packing fractions of polydisperse hard 
spheres [Farr and Groot, J. Chem. Phys. 133, 244104 (2009)]
to the case of lognormal distributions of sphere sizes and mixtures
of such populations. We show that the results compare well
to two much slower algorithms for directly simulating spheres 
in three dimensions, and show that the algorithm is fast enough to tackle
inverse problems in particle packing: designing size distributions to meet
required criteria. The one-dimensional method used in this paper is 
implemented as a computer code in the C programming language, available at
http://sourceforge.net/projects/spherepack1d/ under the terms of the
GNU general public licence (version 2).
\end{abstract}

\maketitle

\section{Introduction}
In granular and mesoscopic systems, various material properties
depend on the close packed volume fraction of the constituent particles. For
example, in the Krieger-Dougherty \cite{KD} relation 
\begin{equation}\label{KrD}
\eta_{r}=(1-\phi/\phi_{\max})^{-[\eta]\phi_{\max}},
\end{equation}
used for estimating the viscosity of
a suspension of hard particles in a Newtonian solvent [where
$\eta_{r}$ is the viscosity relative to that of the solvent, $\phi$ the
volume fraction of the particles and $[\eta]$ a number (equal to $2.5$ for 
spheres)], the viscosity
is predicted to diverge at the packing fraction $\phi_{\max}$.
The value of $\phi_{\max}$
may correspond to a random arrangement at low shear rates
or an aligned `string phase' at high shear rates \cite{Barnes,Ackerson},
but in either case, Eq.\ (\ref{KrD}) implies that this quantity
influences the viscosity over the whole range of volume fractions. 
On the other hand, deformable particles may be packed above 
the Kreiger-Dougherty $\phi_{\max}$,
and their material properties, such as yield stress 
\cite{Mason,Foudazi}, can be deduced from how far above close packing the 
system lies.

For many colloidal and granular systems, the constituent particles do
not form regular, crystalline arrays, but instead are rather randomly
arranged when a jammed state is reached, which represents a close
packed arrangement.
The concept of random close packing (`RCP') was first clearly described for
monodisperse smooth hard spheres by Bernal and Mason \cite{Bernal}, and the 
packing of smooth 
spheres remains an important approximation for less ideal systems.

For the monodisperse case, there has been controversy over the definition
(and even existence \cite{Torquato})
of RCP, as crystallization to a face centred cubic
arrangement \cite{Kepler,Hales} is possible when sufficient opportunity to
explore the configuration space is allowed. Theoretical work on random
jammed states \cite{Silbert} has clarified these issues, but the simplest
evidence for a well-defined RCP state is that different packing algorithms
generally converge to statistically very similar configurations and packing
fractions. One can therefore define RCP operationally, as the outcome
of such a packing algorithm. Various algorithms have been explored:
Conceptually the simplest is the Lubachevski-Stillinger (`LS') 
algorithm \cite{LuSt}
in which spheres at a low volume fraction are placed 
in a box with periodic boundary conditions, by random sequential 
addition. They are then given random initial velocities and permitted 
to move and collide elastically while their radii grow at a rate 
proportional to their initial radius, until a jammed 
state is reached. This algorithm takes three input parameters: the number 
of spheres $N_s$, the initial volume fraction $\phi_{\rm init}$
and the ratio $\delta$ of the radial growth rate to the initial particle size.
For large $N_s$, the final packing fraction is only very weakly 
dependent on $\delta$ and $\phi_{\rm init}$. Usually
fairly large values (around $\delta=0.1$) are chosen, to avoid local
crystalline regions. Even with efficient methods for identifying 
neighbours however, the LS algorithm
converges rather slowly to the jammed RCP state because of the diverging
number of collisions as this point is approached.

Other authors have therefore
modified the dissipative particle dynamics \cite{Warren} method
and simulated smooth, soft (Hertzian) spheres, with radial dissipative forces. 
In the limit of zero confining pressure, these also behave as hard spheres
and give extremely similar results to the LS algorithm,
although the amount of radial dissipation (or equivalently the particle 
size) does have a very weak effect on the final RCP volume fraction \cite{FG}.

In moving toward more realistic systems, there are three constraints
in the above-mentioned models
which one can imagine removing: the smoothness of the particles
(that is to say lack of sliding friction), their spherical shape, and 
monodispersity.

We note in passing that
monodisperse hard spheres, but with the addition of sliding friction, 
have been considered in the literature, and this leads to a family of 
randomly packed states \cite{Song}, with 
random close packing (applying to smooth spheres) and random loose packing 
(highly frictional spheres) being the extreme ends of this spectrum.
Corresponding packing fractions are in the range $0.64$ to $0.53$.
We also note that
random close packing of non-spherical, but smooth particles have also been 
extensively studied;
for example in Ref \cite{Delaney}, different smooth superelliposids are taken
as the objects to be packed.

However, the work reported here will cover only the case of polydisperse
smooth spheres. A certain amount of theoretical effort has been devoted to
this area, notably Refs. \cite{Ouchiyama,Biazzo,Danisch}. The
last of these appears to provide a flexible approximation scheme
that could be applied to fairly general size distributions; although
the authors note that for bidisperse sphere size ratios greater than 2, the 
accuracy declines. Despite these advances, all the theoretical approaches are
to some extent heuristic, requiring comparison to numerical data. 
Therefore the most obvious route forward, which is
to generalize the numerical packing algorithms that were developed
for the monodisperse case, remains necessary. 
In the present paper, the two sets of 3d simulation results we present 
are based on a hard sphere
method (a modification of the LS algorithm \cite{KTS}) and a soft particle
(`SP') algorithm (taken directly from Ref. \cite{FG}).

\begin{figure}
\includegraphics[width=\columnwidth]{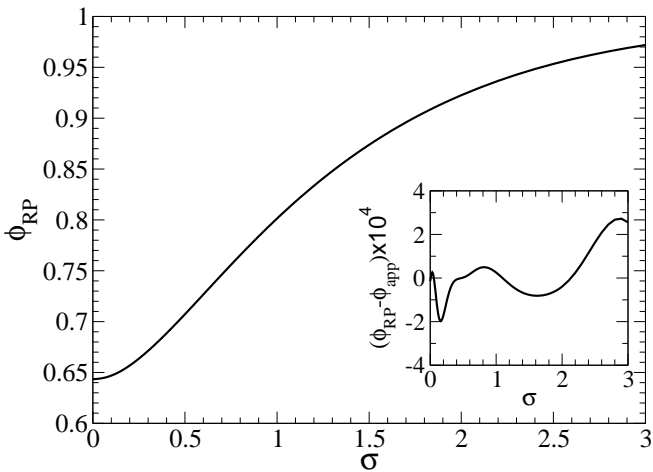}
\caption{\label{lnfig}
Estimate $\phi_{RP}$ (from the RP algorithm of Ref. \cite{FG})
for the random close packed volume fraction of a log-normal
distribution of sphere diameters with width $\sigma$. Insert
shows the error when compared to the fitting function
of Eq. (\ref{lnfit}) in the text.}
\end{figure}

However, all these direct simulation methods are computationally rather 
expensive, typically taking hours or days to obtain high quality results.
Not only is it inconvenient to have to bring to bear a complex and expensive
algorithm when one may only be interested in the random packing of a relatively
simple size distribution, but the slow time for solution makes solving inverse
problems infeasible. By an `inverse problem' we mean searching for a size
distribution which satisfies certain packing criteria, such as finding
the largest RCP volume fraction given a fixed minimum and maximum size for the
particles, or other problems of a similar nature.

Recently however, a quick and apparently quite
accurate algorithm \cite{FG} has been described which attempts to approximate
the random close packing fraction of any distribution of sphere sizes,
by mapping the problem onto a one dimensional system of rods. This can 
allow the RCP volume fraction to be obtained in around one second
(see table \ref{times}),
and therefore makes routine evaluation of these numbers relatively easy.
However, some care is required to implement the algorithm for general
distributions of sphere sizes, and no reference implementation code has 
hitherto been published.

This paper therefore aims to demonstrate that this one dimensional
`rod packing' (RP)
algorithm can be implemented efficiently for typically encountered
sphere size distributions, and also to compare the results to the more
traditional direct simulation approaches above for calculating RCP
volume fractions.

\section{Log-normal size distributions}
\subsection{Analysing experimental data}
Consider a distribution of sphere sizes. Let the number-weighted
distribution of diameters be given by $P_{3d}(D)$, so that the fraction
of the number of spheres with diameters between $D$ and $D+{\rm d}D$
is $P_{3d}(D){\rm d}D$. The volume-weighted distribution of diameters
will then be $P_{\rm vol}(D)\propto D^{3}P_{3d}(D)$, 
while the surface- and diameter-weighted distributions 
will be respectively $P_{\rm surf}(D)\propto D^{2}P_{3d}(D)$ and 
$P_{\rm diam}(D)\propto DP_{3d}(D)$.

For any such number-weighted size distribution $P_{3d}(D)$, one defines
an $m$'th moment by
\begin{equation}\label{moment}
\mu_{m}\equiv\int_{0}^{\infty}D^{m}P_{3d}(D){\rm d}D.
\end{equation}
It is often the case that the the volume-weighted mean diameter $d_{4,3}$ 
and the surface-weighted
mean diameter $d_{3,2}$ are experimentally accessible. They are defined
in terms of the moments via:
\begin{eqnarray}
d_{4,3} &\equiv& \mu_{4}/\mu_{3} \label{d43} \\
d_{3,2} &\equiv& \mu_{3}/\mu_{2} \label{d32}.
\end{eqnarray}

In studies of emulsions \cite{Rajagopal,Hollingsworth}
it is frequently found that the volume-weighted size distribution
of droplets is log-normal, and this can also be
a good approximation for granular
materials, such as sediments \cite{Rosin,Kondolf}.
In general, if $P_{\rm vol}(D)$ is log-normal with a `width' $\sigma$, it 
will have the form: 
\begin{equation}\label{Pvol}
P_{\rm vol}(D)=\frac{1}{D\sigma\sqrt{2\pi}}\exp
\left\{-\frac{\left[\ln(D/D_{0,{\rm vol}})\right]^{2}}{2\sigma^{2}}
\right\},
\end{equation}
where $D_{0,{\rm vol}}$ is a reference diameter setting the scale.
Performing the integrals of Eq. (\ref{d43}) and (\ref{d32}),
we see that 
\begin{equation}
D_{0,{\rm vol}}=(d_{3,2}d_{4,3})^{1/2}.
\end{equation}

We note in passing that one could alternatively define 
a log-normal distribution with particle 
volume, rather than diameter, as the independent variable;
in which case, for the same physical distribution, the volume-based
lognormal width $\sigma_v$ will be $3\sigma$.

Returning to diameter as the independent variable, in experimental work
it is usual to plot the volume-weighted diameter 
distribution on a logarithmic scale, showing the fraction of the
spheres' volume per decade of diameter. If we define $x$ as the
base ten logarithm of the diameter measured in meters (so $x$ counts
the number of decades)
\begin{eqnarray}
x &\equiv& \log_{10}(D/{\rm m}), \\
x_{0} &\equiv& \log_{10}\left( D_{0,{\rm vol}}/{\rm m}\right),
\end{eqnarray}
then the distribution by decade corresponding to $P_{\rm vol}(D)$ is 
\begin{equation}
P_{\rm vol}^{\rm dec}(x)\equiv \frac{dD}{dx}P_{\rm vol}(D)= 
\frac{\ln(10)}{\sigma\sqrt{2\pi}}
\exp\left[-\frac{(x-x_{0})^{2}}{2(\sigma/\ln(10))^{2}}
\right].
\end{equation}
We see that $P_{\rm vol}^{\rm dec}(x)$ has a simple normal distribution
in $x$, and the full width (in decades) 
at half maximum is very close to $\sigma$ itself 
(more precisely $1.023\sigma$).

\begin{figure}
\includegraphics[width=\columnwidth]{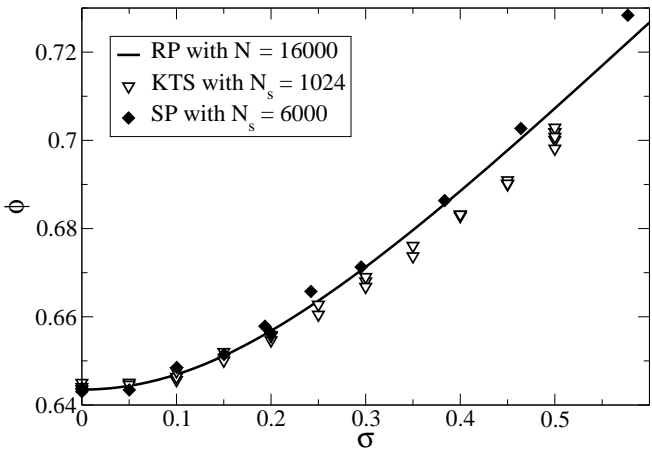}
\caption{\label{lognormal_compare}
Predicted random close packed volume fraction $\phi$ for a collection of
spheres with a lognormal distribution of diameters of width $\sigma$.
Bold curve shoes results from the rod-packing algorithm, with $N=16000$ rods.
Open triangles show results from the hard-sphere KTS algorithm, with
$N_{s}=1024$ particles, and filled diamonds show results taken directly 
from Ref. \cite{FG}, using a soft particle packing algorithm  
with $N_{s}=6000$ spheres.
}
\end{figure}

\subsection{Weighted distributions}
A little algebra shows that if $P_{\rm vol}(D)$ is log-normally distributed,
then so are the number-, diameter- and surface-weighted distributions.
That is to say they have exactly the same functional form as Eq. (\ref{Pvol}),
with the same width $\sigma$, but different values of the reference
diameter. For example the number-weighted diameter distribution is
\begin{equation}\label{Pnum}
P_{3d}(D)=\frac{1}{D\sigma\sqrt{2\pi}}\exp\left\{-\frac{\left[\ln (D/D_{0})
\right]^{2}}{2\sigma^{2}}\right\},
\end{equation}
from which we deduce
\begin{eqnarray}
\mu_{m} &=& D_{0}^{m}\exp(m^2\sigma^2/2), \\
D_{0}   &=& (d_{3,2}d_{4,3})^{1/2}(d_{3,2}/d_{4,3})^{3}, \label{D0}\\
\sigma  &=& \sqrt{\ln\left(\frac{d_{4,3}}{d_{3,2}}\right)},\label{sig_d4_d3}
\end{eqnarray}
while the volume-, surface- and diameter-weighted distributions have
exactly the same functional form as Eq. (\ref{Pnum}) save for $D_{0}$
being replaced by $(d_{3,2}d_{4,3})^{1/2}(d_{3,2}/d_{4,3})^{q}$ with
$q=0$, $1$ and $2$ respectively.

Eq.\ (\ref{sig_d4_d3}) can often be used to estimate $\sigma$ for a 
real lognormal distribution 
of particle diameters, using experimental sizing data, for example
from light-scattering.

As a general observation, if the RCP 
volume fraction $\phi_{RCP}$ is not affected by equally magnifying all the 
spheres (and the boundary conditions of the system), then for a lognormal 
distribution, $\phi_{RCP}$ will depend only
on $\sigma$, which will be the same whether we measure the number-, surface-
or volume-weighted diameter distribution.
Such magnification-independence appears to hold to a good approximation for
the LS, KTS and SP algorithms, and is true exactly for the one-dimensional
rod packing algorithm described in the methods section below.

\begin{figure}
\includegraphics[width=\columnwidth]{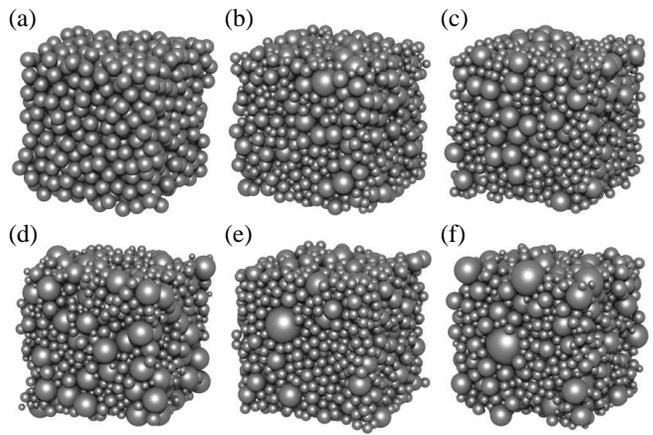}
\caption{\label{size_dist_images}
Random close packed configurations for the size distributions (in order) of 
table \ref{KTStable}, obtained by the KTS algorithm of Ref. \cite{KTS}.
See text for details.
}
\end{figure}

\subsection{Combining several lognormal distributions}
Suppose the distribution of interest is composed of a sum of simpler
distributions, for example monodisperse distributions (so each individual
function $P_{3d}(D)$ is a Dirac delta-function), or lognormal distributions.
Let these normalised number-weighted distributions individually
be the set of functions $\{P_{i}(D)\}$, so that
\begin{equation}\label{total_distribution}
P_{3d}(D)=\sum_{i}a_{i}P_{i}(D)
\end{equation}
where the $a_{i}$'s are the fractions of the total number of particles in
each population, and therefore $\sum a_{i} = 1$.
From a practical point of view, one rarely knows the total number of particles
in a sample of material: more commonly, the occluded volume $v_{i}$ of that 
fraction would be known. This is the total volume occupied by the 
particles themselves, or the volume they would displace in an
Archimedean sense \cite{Arch} were they submerged in a liquid in which they 
were insoluble. If the total mass of this population is $m_{i}$, and the
particles were all made from a material of density $\rho_{i}$, then
$v_{i}=m_{i}/\rho_{i}$. 

We therefore see that the total distribution
of Eq.\ (\ref{total_distribution}) can be constructed from this more
readily available information, because the 
coefficients $a_{i}$ are related
to the normalized occluded volumes $v_{i}$ of the different populations through
\begin{equation}\label{a}
a_{i}=(v_{i}/\mu_{3;i})\left[
\sum_{j}(v_{j}/\mu_{3;j})\right]^{-1} ,
\end{equation}
where
\begin{equation}\label{mu}
\mu_{3;i}\equiv\int_{D=0}^{\infty}D^{3}P_{i}{\rm d}D
\end{equation}
is the third moment of the relevant number-weighted size distribution.

\section{Methods for 3d simulation of sphere packing}
\subsection{Hard spheres}
The LS algorithm has been generalized to polydisperse spheres
by Kansall, Torquato and Stillinger (`KTS') \cite{KTS}, who
apply it to bidisperse packings. To do this, one needs to make some choices
as to how to perform the simulation: In Ref. \cite{KTS} (and the simulations 
here using the same algorithm), the spheres are
all chosen to have equal mass, and furthermore the growth of the radii
must be taken into account at the moment of collision in order
to have a coefficient of restitution of unity. This latter leads
to an increase in kinetic energy of the system at each collision, so
all the velocities are then renormalized, to keep the energy constant.
The radii $\{r_{i}\}$ in this algorithm
are chosen to increase in time according to the relation
\begin{equation}\label{KTS_growth}
r_{i}(t)=(1+t\delta)r_{0;i},
\end{equation}
where $\{ r_{0;i} \}$ are the initial radii. Eq. (\ref{KTS_growth})
has the property that the size distribution remains the same throughout 
the simulation, apart from a uniform magnification.

\subsection{Soft spheres}
The soft particle (`SP') approach 
has also been applied to a range of polydisperse cases in Ref. \cite{FG},
and due to the increased efficiency of this method, allows many thousands 
of spheres to be simulated. For the purposes of this paper, we 
simply quote results directly from Ref. \cite{FG}.

\section{Methods for the rod-packing algorithm}
\subsection{The rod packing model and application to a lognormal distribution}
The 1d algorithm for predicting the random close packing fraction, described
in Ref. \cite{FG}, starts by constructing a normalized distribution
$P_{1d}(L)$ of rod lengths $L$ from any number-weighted diameter distribution
$P_{3d}(D)$. This function is also number-weighted, so that
$P_{1d}(L){\rm d}L$ is the number fraction of rods with lengths between
$L$ and $L+{\rm d}L$. The construction is:
\begin{equation}\label{p1}
P_{1d}(L)=\frac{2L\int_{L}^{\infty} P_{3d}(D){\rm d}D}{
\int_{0}^{\infty}D^{2}P_{3d}(D){\rm d}D}.
\end{equation}

The prediction for the RCP fraction from Ref. \cite{FG} consists in taking 
a collection of rods, with lengths drawn from the distribution $P_{1d}(L)$
and placing them sequentially on a line in the manner described below,
starting from the longest rod,
then the next longest and so on. The rods are not positioned so as to touch
one another, but are instead placed so that there is a gap between any
pair of rods which is at least a fraction $f=0.7654$ of the shorter of the two.
Placement consists in 
repeatedly inserting the longest remaining rod into the largest available
gap, which might require expanding that gap (displacing all rods to the right
of the gap an equal amount to the right) just enough
to insert the new rod while ensuring that the above gap criterion holds 
between every rod pair. In the
case of ambiguity in the placement (for example placing a very small
rod into a large gap, so that many positions are available without
moving the other rods),
we choose the position of such a rod to be at the leftmost end of its possible
positions. In this process we also maintain periodic boundary
conditions in one dimension, so that `expanding a gap' 
also involves increasing the length of the 1d periodic image. 

At the end
of this process (when the smallest rod has been inserted), the rods will
occupy a length fraction $\phi_{RP}$ on the line, and this is the
rod-packing estimate for 
the actual RCP volume fraction of the original
spheres in space. A more detailed description is given in Ref. \cite{FG}.

If we are dealing with a lognormal distribution of spheres, we see
from Eqs. (\ref{Pnum}) and (\ref{p1}), that 
\begin{equation}\label{P1Dratio}
P_{1d}(L)=2L\frac{{\cal I}_{-1}(L)}{{\cal I}_{1}(0)},
\end{equation}
where
\begin{eqnarray}\label{calI}
 {\cal I}_{n}(L) &\equiv& \int_{L}^{\infty}D^{n}
\exp\left\{-\frac{\left[\ln (D/D_{0})
\right]^{2}}{2\sigma^{2}}\right\}{\rm d}D \nonumber \\
   &=& D_{0}^{n+1}\sigma
     \sqrt{\frac{\pi}{2}}e^{(n+1)^{2}\sigma^{2}/2} \nonumber \\
   &\ &\times{\rm erfc}\left[ \frac{\ln (L/D_{0})}{\sigma\sqrt{2}}
-\frac{(n+1)\sigma}{\sqrt{2}}\right],\label{In}
\end{eqnarray}
so that $\mu_{m}\equiv {\cal I}_{m-1}(0)/(\sigma\sqrt{2\pi})$. 
In Eq. (\ref{calI})
\begin{equation}
 {\rm erfc}(x)\equiv\frac{2}{\sqrt{\pi}}\int_{x}^{\infty}
\exp(-t^{2}){\rm d}t
\end{equation}
is the complement of the error function ${\rm erf}(x)$.
Furthermore, using Eqs. (\ref{P1Dratio}) and (\ref{In}), we see finally that
\begin{equation}\label{P1D}
P_{1d}(L)=\frac{Le^{-2\sigma^{2}}}{D_{0}^{2}}
 {\rm erfc}\left[ \frac{\ln (L/D_{0})}{\sigma\sqrt{2}}
\right].
\end{equation}

We now use the RP theory of 
Ref. \cite{FG} to predict the random close packing fraction for lognormal 
distributions of sphere diameters, using different values of $\sigma$. 
We first establish a method to efficiently construct the collection of 
rod lengths needed for the packing algorithm, which is applicable to 
mixtures of (potentially wide) lognormal distributions, as well as the 
single lognormal distribution dealt with immediately below.

\subsection{Rod lengths and analytic fit for a single lognormal distribution}
The RP algorithm requires a sample of rods to be be drawn uniformly from
the distribution $P_{1d}(L)$. This could be done randomly, but a simple 
deterministic method is to find the cumulative distribution function
\begin{equation}\label{cdf}
F(L)\equiv \int_{0}^{L}P_{1d}(L'){\rm d}L',
\end{equation}
and then a set of $N$ rods $\{ L_{i}\}$ with $1\le i\le N$
with strictly non-increasing lengths can be
constructed using the inverse of this function, via
\begin{equation}\label{collection}
L_{i}=F^{-1}\left(\frac{2N-2i+1}{2N}\right).
\end{equation}
Calculating the inverse of a monotonic function can be done relatively 
efficiently using a binary search algorithm \cite{Cormen}, provided $F(L)$ 
can be evaluated quickly. In order to obtain
an explicit form for the function $F(L)$ of Eq. (\ref{cdf}), we use
Eq. (\ref{p1}) and reverse the order of the integrations to obtain 
the identity
\begin{eqnarray}
\mu_{2}F(L) &=& \int_{L'=0}^{L} 
\left[2L'\int_{D=L'}^{\infty}P_{3d}(D){\rm d}D\right] {\rm d}L'\nonumber \\
   &\equiv&
\int_{D=0}^{L}D^{2}P_{3d}(D){\rm d}D \nonumber \\
   &\ & +L^{2}\int_{D=L}^{\infty}
P_{3d}(D){\rm d}D.\label{Fint}
\end{eqnarray}

The integrals of Eq. (\ref{Fint}) can be performed analytically for 
the lognormal distribution of Eq. (\ref{Pnum}) to obtain
\begin{eqnarray}
F(L) &=& 1+\left[L^{2}{\cal I}_{-1}(L)-{\cal I}_{1}(L)\right]
/{\cal I}_{1}(0) \nonumber \\
     &=& 1+\frac{L^{2}}{2D_{0}^{2}}e^{-2\sigma^{2}}
 {\rm erfc}\left[
\frac{\ln(L/D_{0})}{\sigma\sqrt{2}}
\right] \nonumber \\
     &\ & -\frac{1}{2}{\rm erfc}\left[
\frac{\ln(L/D_{0})}{\sigma\sqrt{2}}-\frac{2\sigma}{\sqrt{2}}
\right].  \label{intcdf}
\end{eqnarray}
Eqs. (\ref{intcdf}) and (\ref{collection}) thus provide a method
to construct a collection of rods for the subsequent packing algorithm.

Using a range of values of $\sigma\in(0,3)$, samples of
$N=64000$ rods were chosen from the 
distribution $P_{1d}(L)$, and
these were processed using the RP algorithm of Ref. \cite{FG} to provide
predictions for the RCP volume fractions of these sphere packings. These
rod-packing predictions are denoted $\phi_{RP}(\sigma)$. 
The results are shown in figure \ref{lnfig} and we find that the 
predicted RCP fraction for the spheres can be accurately approximated 
by the arbitrarily constructed, analytic expression
\begin{eqnarray}
\phi_{\rm app}(\sigma) =
1-0.57e^{-\sigma}+0.2135e^{-0.57\sigma/0.2135} \nonumber \\
 + 0.0019\left\{
\cos\left[ 2\pi\left( 1-e^{-0.75\sigma^{0.7}-0.025\sigma^{4}}
\right)\right]-1\right\}.
\label{lnfit}
\end{eqnarray}

Table \ref{times} shows the some example calculation times for the RP
algorithm, using a 3.2GHz Intel Pentium processor and various values
for $N$ and $\sigma$. It is difficult to compare directly the relative
speed of the algorithm here and that in Ref. \cite{FG}, since neither
have been fully optimised for speed. Nevertheless, analytically
performing the integrals and using a binary search to find the rod lengths
does have definite advantages: calculation times of $6541{\rm ms}$ and
$10898{\rm ms}$ were found for the conditions of lines $2$ and $3$ in 
table \ref{times} using the corresponding code from Ref. \cite{FG}.
It is to be anticipated that the advantage of analytic integration over
numerical would only increase as the distributions become wider.

\begin{table}
\caption{\label{times}
Simulation times $t$ in milliseconds for the RP algorithm applied to
a lognormal distribution of spheres, implemented
on a 3.2GHz Intel Pentium processor for various values of rod number $N$
and width $\sigma$. The predicted RCP volume fraction is $\phi_{RP}$.
}
\begin{ruledtabular}
\begin{tabular}{cccc}
$\sigma$ & $N$ & $\phi_{RP}$ & $t$/ms  \\
\hline
    $0.0$  & $16000$ & $0.643485$ & $40$    \\
    $0.5$  & $16000$ & $0.707259$ & $479$    \\
    $1.0$  & $16000$ & $0.801339$ & $507$    \\
    $0.0$  & $64000$ & $0.643485$ & $331$    \\
    $0.5$  & $64000$ & $0.707262$ & $2151$    \\
    $1.0$  & $64000$ & $0.801368$ & $2644$    \\
\end{tabular}
\end{ruledtabular}
\end{table}

\begin{figure}
\includegraphics[width=\columnwidth]{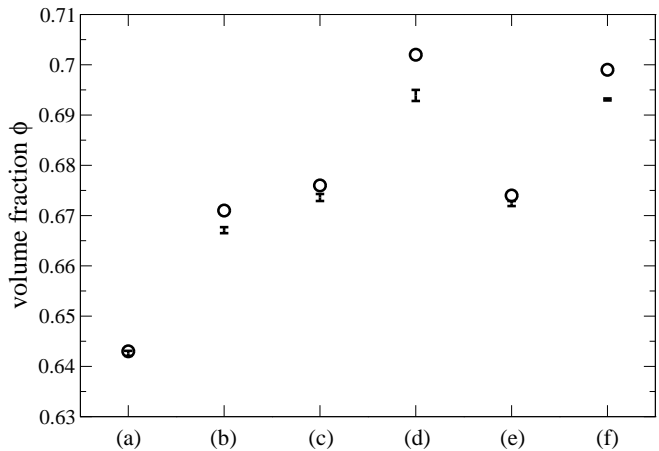}
\caption{\label{distributions_compare}
Data from table \ref{KTStable}, showing a comparison of predictions for
random close packed volume fraction for the sphere size distributions
in table \ref{KTStable}. Data are shown for the
KTS algorithm \cite{KTS} (vertical bars)  and
rod-packing algorithm RP \cite{FG} (open circles).
}
\end{figure}

\subsection{Rod lengths for a sum of monodisperse populations}
If $P_{3d}(D)$ consists of a 
sum of $\delta$-functions, so the individual populations are monodisperse:
\begin{equation}
P_{3d}(D)=\sum_{i}a_{i}\delta(D-D_{i}),
\end{equation}
with $\{ D_{i}\}$ being ordered such that $D_{i}>D_{j}$ if $i>j$, then
from Eqs. (\ref{mu}) and (\ref{a})
\begin{equation}
a_{i}=(v_{i}/D_{i}^{3})\left[
\sum_{j}(v_{j}/D_{j}^{3})\right]^{-1} .
\end{equation}
Eq. (\ref{Fint}) then becomes
\begin{equation}
F(L)=\left[
\sum_{i:D_{i}\le L}\left(a_{i}D_{i}^{2}\right)+
\sum_{i:D_{i}> L}\left(a_{i}L^{2}\right)
\right]\left/
\sum_{i}\left(a_{i}D_{i}^{2}\right)\right. ,
\end{equation}
and the rod lengths can be calculated directly from Eq. (\ref{collection}).

\subsection{Rod lengths for a sum of lognormal populations}
Suppose instead we have a mixture of lognormal distributions, with normalized
number-weighted distributions of diameters $\{P_{i}\}$, and suppose that
for each of these, we know the 
volume-weighted and surface-weighted mean diameters
$d_{4,3;i}$ and $d_{3,2;i}$ and also the occluded volume $v_{i}$ of
each population (a plausible scenario if the actual distribution is made by
physically mixing different monomodal fractions). We define 
the (log) width of each population by $\sigma_{i}$, which we see from
Eq. (\ref{sig_d4_d3}) is
\begin{equation}
\sigma_{i}\equiv \sqrt{\ln(d_{4,3;i}/d_{3,2;i})}.
\end{equation}
From Eq. (\ref{Pnum}) and (\ref{D0}) we find
\begin{eqnarray}
P_{i}(D) &=& \frac{1}{D\sigma_{i}\sqrt{2\pi}}
\exp\left\{-
\frac{[\ln(e^{7\sigma_{i}^{2}/2}D/d_{4,3;i})]^{2}}{2\sigma_{i}^{2}}
\right\} \\
\mu_{3;i} &=& (d_{4,3;i})^{3}e^{-6\sigma_{i}^{2}},
\end{eqnarray}
from which [using Eq.(\ref{a})] we can calculate the $a_{i}$'s. Then 
from Eq. (\ref{Fint}) we find 
\begin{equation}\label{part1}
F(L)=\nu^{-1}\sum_{i}\left\{ a_{i}\left[A_{i}(d_{4,3;i})^{2}+B_{i}L^{2}
\right]\right\} 
\end{equation}
where
\begin{eqnarray}
\nu    &\equiv& \sum_{i}a_{i}(d_{4,3;i})^{2}e^{-5\sigma_{i}^{2}}, \\
A_{i}  &\equiv& \frac{e^{-5\sigma_{i}^{2}}}{2}
\left\{ 2-{\rm erfc}\left[
\frac{\ln (e^{7\sigma_{i}^{2}/2}L/d_{4,3;i})}{\sigma_{i}\sqrt{2}}
-\frac{2\sigma_{i}}{\sqrt{2}}\right]\right\} \\
B_{i}  &\equiv& \frac{1}{2}{\rm erfc}\left[
\frac{\ln (e^{7\sigma_{i}^{2}/2}L/d_{4,3;i})}{\sigma_{i}\sqrt{2}}
\right].\label{part4}
\end{eqnarray}
Collectively, Eqs. (\ref{part1}) to (\ref{part4}) provide an
efficient method to calculate the cumulative distribution $F(L)$ to which
Eq. (\ref{collection}) is applied. Efficiency comes from explicitly 
evaluating the double integrals in Eq.\ (\ref{Fint}) and (to a 
lesser degree) eliminating the sorting of rods which is implicit in
the description of the algorithm in Ref. \cite{FG}.

\begin{figure}
\includegraphics[width=\columnwidth]{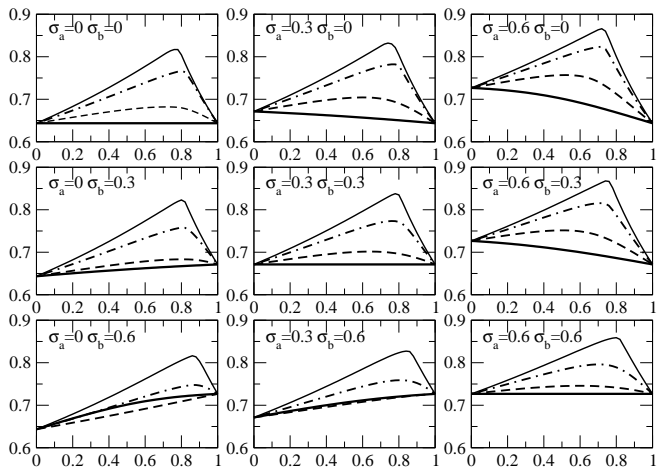}
\caption{\label{bim_logn}
Plots of predicted RCP fraction from the RP algorithm, for a mixture of 
two lognormal populations
of spheres. The two populations are denoted `$a$' and `$b$', and have different
widths $\sigma_{a}$ and $\sigma_{b}$, as shown in the figure. In each panel
the horizontal axis shows the proportion (by occluded volume) of the
population `$b$' in the mixture: at the left hand end, the system is entirely
population `$a$', and at the right hand end entirely population `$b$'. 
Each panel contains four curves, corresponding to different values
of $R\equiv d_{4,3;b}/d_{4,3;a}$: $R=1$ (heavy solid line), $2$ (dashed line),
$4$ (dash-dot line) and $8$ (thin solid line).
}
\end{figure}

\section{Results and discussion}
Figure \ref{lognormal_compare} shows results for the random close
packed volume fraction of a single lognormal size distribution,
as a function of the distribution width $\sigma$. We see that the 
rod-packing algorithm (RP) gives good agreement with the KTS results
using $N_{s}=1024$ spheres, and excellent agreement with the soft particle
packing algorithm SP, using $N_{s}=6000$ spheres. The results from the KTS
algorithm, with the smaller number of spheres, tend to be slightly below the
RP results and those from the SP algorithm. The SP results use a larger number 
of spheres, and are therefore probably more accurate than the KTS results.

As a further comparison, we also look at mixtures of two lognormal
size distributions; a family of problems which appears to be little
studied in the literature. The two lognormal populations `$a$' and `$b$' 
which are combined have widths $\sigma_a$ and $\sigma_b$ respectively, 
and volume-weighted mean diameters $d_{4,3;a}$ and $d_{4,3;b}$.
We define a size ratio
\begin{equation}
R\equiv d_{4,3;b}/d_{4,3;a},
\end{equation}
and define $w$ to be the ratio of the occluded volume of the `$b$' population
to that of both populations together. Thus, if both populations were
made of the same material, then $w$ is the fraction of the mass of
the particles that is due to the `$b$' population.

The results are shown in table \ref{KTStable} and figure 
\ref{distributions_compare}, and some of the close-packed configurations
from the KTS algorithm are shown in figure \ref{size_dist_images}.
Again, we see good agreement between the results of the rod-packing algorithm
and the predictions from the KTS algorithm, but once more, the KTS results are
seen to be in general slightly below the RP results

\begin{table}
\caption{\label{KTStable}
Predicted random close packed volume fractions for mixtures of
two lognormal populations of sphere sizes, denoted `$a$' and `$b$'.
$R$ is defined as $d_{4,3;b}/d_{3,4;a}$, and $w$ is the proportion 
(by occluded volume) 
of the `$b$' population in the mixture.
Results are shown for the KTS \cite{KTS} algorithm ($\phi_{KTS}$, with
values based on three repeats) and the rod-packing \cite{FG}
algorithm ($\phi_{RP}$). The initial volume fraction in the KTS
algorithm is $\phi_{\rm init}$.
}
\begin{ruledtabular}
\begin{tabular}{cccccccc}
$N_{s}$ & $R$ & $\sigma_a$ & $\sigma_b$ & $w$ & $\phi_{\rm init}$ & 
   $\phi_{KTS}$ & $\phi_{RP}$  \\
\hline
    1024  & $-$ & $0$   & $-$   & $-$   & $0.2$  & $0.6426 \pm 0.0005$ & $0.643$ \\
    2048  & $-$ & $0.3$ & $-$   & $-$   & $0.15$ & $0.6671 \pm 0.0006$ & $0.671$ \\
    2048  & $2$ & $0$   & $0$   & $0.5$ & $0.15$ & $0.6736 \pm 0.0007$ & $0.676$ \\
    2048  & $2$ & $0.3$ & $0$   & $0.5$ & $0.15$ & $0.6939 \pm 0.0011$ & $0.702$ \\
    2048  & $2$ & $0$   & $0.3$ & $0.5$ & $0.15$ & $0.6724 \pm 0.0005$ & $0.674$ \\
    2048  & $2$ & $0.3$ & $0.3$ & $0.5$ & $0.15$ & $0.6931 \pm 0.0002$ & $0.699$ \\
\end{tabular}
\end{ruledtabular}
\end{table}

Having provided evidence that the one dimensional RP model gives good
agreement with the more traditional simulation methods, for a
fairly broad range of sphere size distributions, we now demonstrate that 
it can be used to map out parameter spaces and make interesting predictions 
beyond the scope of direct simulation approaches.

For example, consider the space consisting of all possible mixtures of any 
two lognormal distributions of sphere sizes. Figure \ref{bim_logn} shows 
plots of predicted random close packing fractions
for this parameter space. The lower left hand panel, which 
covers $\sigma_{a}=0$ and $\sigma_{b}=0.6$, shows the interesting
phenomenon that when one distribution is much wider than the other, and 
for certain ratios of occluded volumes of the two populations,
the lowest packing fraction may not be achieved for equal values of $d_{4,3}$.
This is a consequence of the lognormal distribution becoming markedly
skewed when it is broad.

\begin{figure}
\includegraphics[width=\columnwidth]{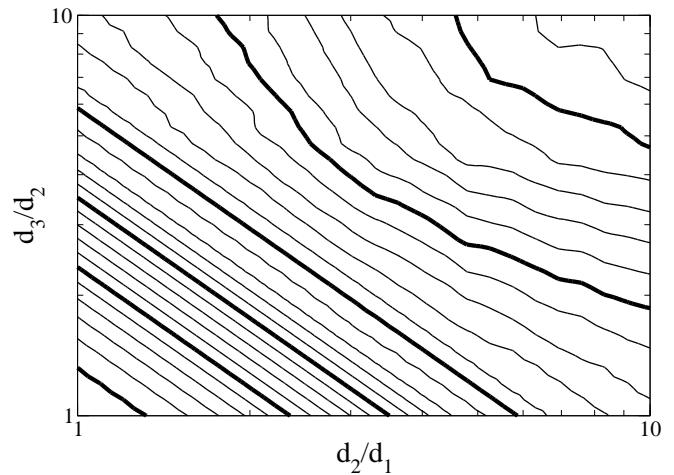}
\caption{\label{tri}
Predicted optimised RCP fraction of tridisperse sphere mixtures
using the RP algorithm.
The individual populations have diameters $d_{1}$, $d_{2}$ and $d_{3}$. 
For each such choice, the contour plot shows the maximum RCP fraction 
when the relative amounts of each of the populations have been optimized
(a two dimensional space of possibilities). Contours are at intervals
of $0.01$, with the bold contours being
$\phi_{RP}=0.65$ (at bottom left), $0.7$, $0.75$, $0.8$, $0.85$ and $0.9$ 
(near the top right).
}
\end{figure}

With the increased speed of the RP algorithm, it becomes practical
not just to make predictions for packing fractions, but to search over
moderately large spaces of distribution functions in order to find optima.
Figure \ref{tri} shows the predicted {\em maximum} RCP fraction for 
mixtures of tridisperse spheres (three populations of monosize spheres).
Each point on the contour plot corresponds
to a fixed pair of ratios of sizes, and represents an optimisation over
all possible ratios of occluded volume (a two-dimensional composition space).
Such optimisation problems are difficult to approach experimentally
or computationally using more tradition packing algorithms, although
the results can be verified by either approach. Therefore an
accelerated algorithm can in this case widen the class of problems
amenable to solution.

Because the algorithms used here require some patience to construct in 
computer code, a reference implementation, written in the `C' programming
language \cite{C} has been written, and made available under 
the open source GNU `general public licence' (GPL version 2 \cite{GPL}) 
at the website indicated in Ref. \cite{SF}.

\section{Acknowledgements}
The research leading to these results has received funding from
the European Community's Seventh Framework Programme [FP7/2007-2013]
under grant agreement 214015. The author would like to thank an
anonymous referee for very detailed and helpful criticism.

\pagebreak\ 

\pagebreak

\section{Appendix: Code for the Rod-Packing (`RP') algorithm}
This section contains the C code from Ref. \cite{SF}. If you wish to 
use the code from the present document, it is most easily obtained 
from the \LaTeX\ file used to generate the file you are now reading.
The program should be fairly self-explanatory. To build and
run it, put the source code
in a file called {\tt "spherepack1d.c"}, then compile it, for example using
the GNU compiler: {\tt "gcc spherepack1d.c -o spherepack1d -lm -O3"}.
Finally, run it from the command line with the 
help option: {\tt "spherepack1d -h"} and it will provide instructions.

\begin{tiny}
\begin{verbatim}

/*-------------------------------------------------------------*
*                        spherepack1d                          *
*                                                              *
*  A program to estimate the maximum packing fraction of       *
*  hard spheers with various distributions of diameters,       *
*  based on an algorithm published in:                         *
*  Journal of Chemical Physics, 131, 244104 (2009).            *
*                                                              *
*  Author: Robert Farr.                                        * 
*                                                              * 
*  Released under the GNU General Public Licence, version 2.   *
*                                                              * 
*               version 1.2 (1/6/2013)                         * 
*                                                              *
*  Changes since version 1.0:                                  *
*  - Corrected some errors in the documentation (-h)           *
*  - Error messsages now sent to stderr, not stdout            *
*  - New '-w' option for inputting the log width directly      *
*                                                              *
*  Changes since version 1.1:                                  *
*  - Lines of code are all less than 90 characters (to make    *
*    formatting in ArXiV version neater).                      *
*                                                              *
*--------------------------------------------------------------*/

#include<unistd.h>
#include<stdio.h>
#include<math.h>

#define NMAX (100001)
#define PMAX (10001)

// Lots of global parameters: Sorry, I guess it's a bad fortran habit.
// Here they all are:
// List of lengths (which must be sorted in decreasing order. Note that
// we don't use the [0] entry, to accord better with the notation in the paper.
// This is a bit ugly. Sorry.
double L[NMAX];  
// List of gaps
double g[2*NMAX];
// List of values for the volume weighted mean diameter, (from 0 to p-1):
double d43[PMAX];
// List of values for the surface weighted mean diameter, (from 0 to p-1):
double d32[PMAX];
// List of values for the occluded volume (equal to the mass if the material
// density is always the same) of each of the populations (from 0 to p-1):
double vol[PMAX];
// Set of 'a' parameters, which are the relative number of particles
// present in each of the p different populations. This is calculated
// from the values of 'vol' and 'd43' (and, for the lognormal case, d32).
double a[PMAX];
// Set of sigma parameters, needed for the case where P_{3d} is a
// sum over lognormal distributions.
double sigma[PMAX];
// Normalization parameter, needed when P_{3d} is a sum of delta functions.
double norm;
// Packing 'f' parameter:
double f;
// Number of rods:
int N;
// Number of lognormal distributions or delta functions
int p;
// flag for the -l -d and -w options:
// The value is 0 for '-d', 1 for '-l' and 2 for '-w'
int lflag;

void set_defaults(){
// Default number of rod lengths to use:
   N=16000;
// Default value of 'f' aprameter corresponds to sphere packing:
   f=0.7654;
// Default number of populations of sphere sizes:
   p=1;
// Default is to use delta functions rather than lognormal distributions:
   lflag=0;
}

double erf(double x){
    // Some versions of math.h don't include the error function, so here
    // is some public domain code ( obtained from John.D.Cook's website at
    // http://www.johndcook.com many thanks for an excellent resource!). 
    //
    // The following are the required constants for the calculation:
    double a1 =  0.254829592;
    double a2 = -0.284496736;
    double a3 =  1.421413741;
    double a4 = -1.453152027;
    double a5 =  1.061405429;
    double p  =  0.3275911;

    // Save the sign of x
    int sign = 1;
    if (x < 0)
        sign = -1;
    x = fabs(x);

    // Abramowitz & Stegun formula 7.1.26
    double t = 1.0/(1.0 + p*x);
    double y = 1.0 - (((((a5*t + a4)*t) + a3)*t + a2)*t + a1)*t*exp(-x*x);

    return(sign*y);
}

double erfc(double x){return(1.0-erf(x));}

void get_a_and_other_things(){
//  This calculates the set of 'a' parameters.
//  It also calculates the normalization parameter 'norm'. Also the sigma
//  parameters when the distribution is a sum of lognormal distributions
//  specified by d43 and d32 ('-l' option). In the case where the lognormal 
//  distributions are specified by d43 and sigma ('-w' option), then we
//  have all the information we need, because d32 is not needed here.
//  Nevertheless, we do calculate it, for completeness.
    int i;
    double sum_mu;
    sum_mu=0.0;
    if(lflag==1 || lflag==2){
//    P_{3d} is a sum of lognormal distributions
      if(lflag==1){for(i=0;i<p;i++){sigma[i]=sqrt(log(d43[i]/d32[i]));}}
      if(lflag==2){for(i=0;i<p;i++){d32[i]=d43[i]*exp(-sigma[i]*sigma[i]);}}
      for(i=0;i<p;i++){sum_mu+=exp(6.0*sigma[i]*sigma[i])*vol[i]/
         (d43[i]*d43[i]*d43[i]);}
      for(i=0;i<p;i++){a[i]=exp(6.0*sigma[i]*sigma[i])*vol[i]/
         (d43[i]*d43[i]*d43[i]*sum_mu);}
//    We also need a normalization parameter, which we calculate thus:
      norm=0.0;
      for(i=0;i<p;i++){norm+=a[i]*d43[i]*d43[i]*exp(-5.0*sigma[i]*sigma[i]);}
    } else {
//    P_{3d} is a sum of delta functions
      for(i=0;i<p;i++){sum_mu+=vol[i]/(d43[i]*d43[i]*d43[i]);}
      for(i=0;i<p;i++){a[i]=vol[i]/(d43[i]*d43[i]*d43[i]*sum_mu);}
//    We also need a normalization parameter, which we calculate thus:
      norm=0.0;
      for(i=0;i<p;i++){norm+=a[i]*d43[i]*d43[i];}
    }
}

double intP1d(double x){
//    This function is the integral of the 1d rod length
//    distribution function (which needs to be inverted in order
//    to obtain the rod lengths). It is denoted by 'F' in recent work.
      double F;
      double Ai,Bi,work;
      int i,j;
      F=0.0;
      if(lflag==1 || lflag==2){
//      Size distribution is a sum of lognormal distributions:
        for(i=0;i<p;i++){
          work=log(exp(7.0*sigma[i]*sigma[i]/2.0)*x/d43[i])
             /(sigma[i]*sqrt(2.0));
          Ai=0.5*exp(-5.0*sigma[i]*sigma[i])*(2.0-erfc(work-
             (2.0*sigma[i]/sqrt(2.0))));
          Bi=0.5*erfc(work);
          F+=a[i]*(Ai*d43[i]*d43[i]+Bi*x*x);
        }
        F=F/norm;
      } else {
//      Size distribution is a sum of delta-functions. For efficiency,
//      we demand that these delta functions are arranged in order of
//      increasing size of the constituent particles.
        for(i=0;i<p;i++){
          if(d43[i] <= x){
            F+=a[i]*d43[i]*d43[i];
          } else {
            F+=a[i]*x*x;
          }
        }
        F=F/norm;
      }
      return(F);
}
      
double inverseF(double x){
      // This function finds the inverse of the function intP1d()
      double Lmin,Lmid,Lmax;
      double F;
      int i;
      Lmin=0.0;
      Lmax=10.0;
      while(intP1d(Lmax) <= x){
        Lmax=Lmax*1.5;
      }
      for(i=1;i<=100;i++){
       Lmid=0.5*(Lmin+Lmax);
       F=intP1d(Lmid);
       if(F > x){
         Lmax=Lmid;
       } else {
         Lmin=Lmid;
       }
      }
      return(Lmid);
}

void getL(){
//     This subroutine generates the list {L} of
//     rod lengths in decreasing (or non-increasing)
//     order. Once this subroutine has been called, we are ready to start
//     calculating the maximum packing fraction.
       int iL;
       double x;
//     Then the rod lengths are given by L_{i} where
//     intP1d(L_{i})=(2*N-2*i+1)/(2*N)
       for(iL=1;iL <= N;iL++){
        x=(2.0*N-2.0*iL+1.0)/(2.0*N);
        L[iL]=inverseF(x);
       }
}


double phimax(){
//    This function works out the maximum packing fraction, given
//    a list of N lengths {L} which is DECREASING (or rather non-increasing)
//    and the 'f' parameter. This function therefore assumes that the
//    set {L} is already sorted.
//
//    The index of the biggest and smallest gaps
      int gapfirst,gaplast;
//    Total rod and gap lengths:
      double Ltot,gtot;
//    Some working variables
      int i,j,ig,igg,index;
      double store,gap[3];
      g[1]=f*L[1];
      gapfirst=1;
      gaplast=1;
      if(N == 1){
        fprintf(stderr,"Need N>1\n");
        return(-1.0);
      }
      for(i=2;i <= N;i++){
//     we now destroy the smallest gap, and replace it by two
//     new gaps, called gap(1) and gap(2)
       gapfirst=gapfirst+1;
       gap[1]=f*L[i];
       if((2.0*f+1.0)*L[i] >= g[gapfirst-1]){
//       we end up with two equal sized gaps, because we have shoved apart
//       the largest existing gap, and both sides rest snugly against
//       the ends of the inserted rod:
         gap[2]=f*L[i];
       } else {
//       there is still some wiggle-room at the right hand end,
//       and so this gap is somewhat larger than f*L(i):
         gap[2]=g[gapfirst-1]-(f+1.0)*L[i];
       }
//     OK, now we have to insert each of the two new gaps into the
//     ordered list of (decreasing) gaps that will run from i to 2i-1
       if(i == 2){
//       note, gap[2] may be larger, so we insert it first in the list:
         g[gapfirst]=gap[2];
         gaplast=gapfirst+1;
         g[gaplast]=gap[1];
       } else {
         for(ig=1;ig<=2;ig++){
          if(gap[ig] <= g[gaplast]){
//          insert at the end:
            gaplast=gaplast+1;
            g[gaplast]=gap[ig];
          } else {
            for(igg=gapfirst;igg <= gaplast;igg++){
//           so, we go through the current list of gaps:
             if(g[igg] <= gap[ig]){
               index=igg;
               break;
             }
            }
//          now insert the new gap
            gaplast=gaplast+1;
            for(igg=gaplast;igg >= index+1;igg--){
             g[igg]=g[igg-1];
            }
            g[index]=gap[ig];
          }
         }
       }
      }
//    Calculate the volume fraction:
      Ltot=0.0;
      gtot=0.0;
      for(i=1;i<=N;i++){
       Ltot=Ltot+L[i];
      }
      for(i=gapfirst;i<=gaplast;i++){
       gtot=gtot+g[i];
      }
      return(Ltot/(Ltot+gtot));
}



int main(int argc, char *argv[ ]) {
    int c;
    char *number_string;
    extern char *optarg;
    extern int optind, optopt, opterr;

    int verbose=0;    // Are we in verbose mode?
    int iflag=0;      // flag for the -i option
    int uflag=0;      // flag for the -u option
    int fflag=0;      // flag for the -f option
    int pflag=0;      // flag for the -p option
    int sflag=0;      // flag for the -s option

    int zero_volume_flag;
    int i,j;
//  variable needed for the -s case only:
    double sig;

    set_defaults();

    while ((c = getopt(argc, argv, ":hVdlwuvisp:N:f:")) != -1) {
        switch(c) {
        case 'h':
            printf("\
NAME\n\
       spherepack1d - calculate approximate random close packed volume\n\
                   fraction for a distribution of sphere sizes\n\
\n\
SYNOPSIS\n\
       spherepack1d [-dhilsuvVw][-p integer][-N integer]\n\
       [[integer]]\n\
       [real] [real] [real]\n\
       [real] [real] [real]\n\
       ...\n\
\n\
DESCRIPTION\n\
       The sphere size distribution is specified through the input data\n\
       and the estimated random close packed volume fraction is calculated\n\
       by an algorithm described in [Farr and Groot, J. Chem. Phys.\n\
       volume 131, article no. 244104 (2009)].\n\
\n\
       -d  Input a sequence of delta functions for the size distribution.\n\
           If neither [-d] nor [-l] option is specified, [-d] is assumed\n\
\n\
       -l  Input a sequence of lognormal functions for the size distribution\n\
           by specifying d43, d32 and occluded volume for each. The occluded\n\
           volume is not needed if only one lognormal distribution is given.\n\
\n\
       -w  Input a sequence of lognormal functions for the size distribution\n\
           by specifying d43, log-width (sigma) and occluded volume for each.\n\
           Only sigma is needed if only one lognormal distribution is given.\n\
\n\
       -f  Specify the 'f' parameter in the packing algorithm. This is done\n\
           by inputting an integer, which is divided by 10^4 to arrive at the\n\
           value of 'f'. The default value of f is 0.7654, which would be\n\
           input to this option as the integer 7654\n\
\n\
       -h  Display this help message and exit\n\
\n\
       -N  Specify the number of rods to use in the algorithm.\n\
           Default value is 16000. Allowed range is 2 to 100000.\n\
\n\
       -p  Number of lines of data which the program will expect.\n\
           This is equal to the number of lognormal distributions (-l option)\n\
           or delta functions (-d option; the default) which go to make up\n\
           the sphere size distribution. Default value is 1.\n\
\n\
       -i  Instead of specifying the number p of populations on the command\n\
           line, the number is specified as the first line read in from\n\
           standard input. This is useful if the program is being fed data\n\
           from a file, rather than via standard input.\n\
\n\
       -s  For the case of a single lognormal distribution, gives an answer\n\
           based on a simple analytic approximation to the results of the\n\
           usual algorithm. The advantage is a much faster answer\n\
\n\
       -u  Uniform data input format: the program will always require three\n\
           data items for the [-l] option, and two for the [-d] option,\n\
           even if there is only one population (p=1)\n\
\n\
       -v  Use verbose mode: the input data are explicitly requested.\n\
\n\
       -V  Print version number and exit.\n\
\n\
FORMAT OF INPUT DATA FOR THE [-l] (LOGNORMAL) MODE\n\
        When the program is run, it expects a series of numbers to be\n\
        supplied from the  standard input. For the [-l] mode where\n\
        the sphere size distribution is a sum of lognormal distributions,\n\
        the format is three real numbers specifying each of these\n\
        lognormal distributions. For example, the following\n\
\n\
           ./spherepack1d -l -p 2\n\
             d43_1 d32_1 vol_1\n\
             d43_2 d32_2 vol_2\n\
\n\
        would calculate the close packed volume fraction for a mixture\n\
        of two log-normal distributions (the [-p 2] option), where for the\n\
        first one, the volume weighted mean diameter is d43_1, the surface-\n\
        weighted mean diameter is d32_1 and the occluded volume of the\n\
        first population of spheres is vol_1. The second population is\n\
        specified similarly in the second line. The width sigma of each\n\
        lognormal distribution is calculated in the code from\n\
\n\
           exp(sigma^2)=d43/d32.\n\
\n\
        The occluded volume of each population is the total volume occupied\n\
        by the particles themselves, or the volume they would displace (in an\n\
        Archimedean sense) were they submerged in a liquid in which they\n\
        were insoluble. For the case when all the sphere poulations\n\
        are made from a material with the same density,\n\
        then vol_1 and vol_2 can be replaced by the total mass of each\n\
        population of spheres.\n\
\n\
        If -p is given the value 1 (a single lognormal\n\
        distribution) then vol_1 is omitted (see examples below).\n\
\n\
FORMAT OF DATA INPUT FOR THE [-d] (DELTA-FUNCTION) MODE\n\
        When the program is run, it expects a series of numbers to be\n\
        supplied from the  standard input. For the default mode where\n\
        the sphere size distribution is a sum of monodisperse\n\
        (delta-function) distributions, the format is two real numbers \n\
        specifying the diameter and occluded volume for each of these\n\
        monodisperse populations. For example, the following\n\
\n\
           ./spherepack1d -d -p 3\n\
             diameter_1 vol_1\n\
             diameter_2 vol_2\n\
             diameter_3 vol_3\n\
\n\
        would calculate the close packed volume fraction for a mixture\n\
        of three monodisperse distributions (the [-p 3] option), where for\n\
        the first one, the sphere diameters are all diameter_1, \n\
        and the occluded volume of the the spheres in this population\n\
        is vol_1. As noted above, 'occluded volume' is the volume that\n\
        the spheres would displace if submerged in a liquid in which they\n\
        are not soluble.\n\
\n\
EXAMPLES\n\
\n\
        To calculate the maximum packing fraction of a single lognormal\n\
        distribution with d43=2.0 and d32=0.5, one would type at the\n\
        command prompt the following:\n\
\n\
           ./spherepack1d -l\n\
           2.0 0.5\n\
\n\
        and the program would produce as output 0.829998.\n\
\n\
        To calculate the maximum packing fraction of equal masses of\n\
        two lognormal distributions, one with d43=5.0 and d32=4.0, and\n\
        the other with d43=3.0, d32=2.0, one would type:\n\
\n\
           ./spherepack1d -l -p 2\n\
           5.0 4.0 1.0\n\
           3.0 2.0 1.0\n\
\n\
        To calculate the maximum packing fraction of a mixture of three\n\
        monodisperse populations of spheres, with diameters of\n\
        2.0, 3.0 and 4.0, and relative masses of these populations in\n\
        the ratio 1.5:2.5:3.5, one would type:\n\
\n\
           ./spherepack1d -d -p 3\n\
           2.0 1.5\n\
           3.0 2.5\n\
           4.0 3.5\n\
\n\
        (Note that in this case, the [-d] option does not need to be\n\
        specified, as it is the default).\n\
\n\
        An alternative way to specify the same problem, would be to use\n\
        the [-i] option, so that the number of populations is specified\n\
        as the first line of standard input. Thus the following would\n\
        give exactly the same answer:\n\
\n\
           ./spherepack1d -di\n\
           3\n\
           2.0 1.5\n\
           3.0 2.5\n\
           4.0 3.5\n\
\n\
AUTHOR\n\
        Robert Farr, Unilever Research, Colworth House, Bedford, UK\n\
        and The London Institute for Mathematical Sciences, Mayfair, London.\n\
\n");
            return(0);
            break;
        case 'V':
            printf("Version 1.0 Released 29/02/2012.\n");
            return(0);
            break;
        case 'v':
            printf(">>> Verbose mode <<<\n");
            verbose=1;
            break;
        case 'N':
            number_string = optarg;
            N=atoi(number_string);
            if(N<2 || N>=NMAX){
              fprintf(stderr,"Error: Number of rods needs to be in the");
              fprintf(stderr," range 2 to %i.\n",NMAX-1);
              return(-1);
            }
            if(verbose){printf("Number of rods is %i\n",N);}
            break;
        case 'p':
            pflag=1;
//          If -i was not chosen, then read in the value of p from 
//          the command line:
            number_string = optarg;
            p=atoi(number_string);
            if(p<1 || p>= PMAX){
              fprintf(stderr,"Error: Number of sphere populations needs to");
              fprintf(stderr," be in range 1 to %i.\n",PMAX-1);
              return(-1);
            }
            if(verbose){printf("Number of data lines is %i\n", p);}
            break;
        case 'f':
            fflag=1;
            number_string = optarg;
            f=0.0001*atoi(number_string);
            break;
        case 'i':
            iflag=1;
            break;
        case 's':
            sflag=1;
            break;
        case 'u':
            uflag=1;
            break;
        case 'l':
            lflag=1;
            break;
        case 'w':
            lflag=2;
            break;
        case 'd':
            lflag=0;
            break;
        case ':':
            printf("-%c without integer\n", optopt);
            break;
        case '?':
            printf("unknown arg %c\n", optopt);
            break;
        }
    }

//  Give an error if the user has asked to specify the number
//  of populations in the data entry rather than the command line:
    if(iflag && pflag){
      fprintf(stderr,"Error: -i option has been chosen, which is not");
      fprintf(stderr," compatible with entering\nthe number of populations");
      fprintf(stderr," in the command line.\n");
      return(-1);
    }
//  Deal with the -s option first:
    if(sflag){
      if(p != 1){
        fprintf(stderr,"Error: -s option requires a single population.\n");
        return(-1);
      }
      if(iflag){
        fprintf(stderr,"Error: -s option is not compatible with -i option \n");
        fprintf(stderr,"(it assumes a single population).\n");
        return(-1);
      }
      if(uflag){
        fprintf(stderr,"Error: -s option is not compatible with -u option \n");
        fprintf(stderr,"(there is no need to specify the volume).\n");
        return(-1);
      }
    }
//  Give some more user feedback in the verbose case:
    if(verbose && lflag == 0){
      printf("The sphere size distribution is a sum of delta functions\n");
    }
    if(verbose && (lflag>0)){
      printf("The sphere size distribution is a sum of lognormal");
      printf(" distributions.\n");
    }
    if(verbose && uflag){
      printf("Data format will be the same for p=1 as for p>1.\n");
    }
    if(verbose && iflag){
      printf("The number of populations p is specified as the first line");
      printf(" of\nstandard input, not on the command line.\n");
    }
    if(verbose && fflag){printf("Value of 'f' parameter is %lf\n", f);}

//  OK, now read in the number of populations, if this was not specified
//  in the command line options.
    if(iflag){
      if(verbose){
        if(lflag==1 || lflag==2){
          printf("How many many lognormal populations does the");
          printf(" distribution consist of?\n");
        } else {
          printf("How many many delta-function populations does the");
          printf(" distribution consist of?\n");
        }
      }
      scanf("%d",&p);
      if(p<1 || p>= PMAX){
        fprintf(stderr,"Error: Number of sphere populations needs to");
        fprintf(stderr," be in range 1 to %i.\n",PMAX-1);
        return(-1);
      }
    }

//  Now do the analysis for the of: (lflag==1) lognormal (specify d43 and d32)
//                                  (lflag==2) lognormal (specify sigma)
//                                  (lflag==0) delta-functions
    if(lflag==1){
//    This is the case where the sphere size distributino is a sum
//    of lognormal distributions:
      if(p==1 && uflag==0){
//      Read in the data, which consists of the pairs (d43,d32):
        if(verbose){
          printf("Please input data on the lognormal population, in the");
          printf(" format of a\n  d34 d32 \npair\n",p);
        }
        scanf("%lf %lf",&d43[0],&d32[0]);
        vol[0]=1.0;
        if(d43[0]<=0.0 || d32[0]<=0.0 || d32[0]>d43[0]){
          fprintf(stderr,"Error: must have d43 > d32 > 0.\n");
          return(-1);
        }
//      Now deal with the -s case:
//------This piece of code uses a simple analytic approximation
//      for the case of one lognormal distribution:
        if(sflag){
          if(verbose){printf("\nPredicted maximum packing fraction:\n");}
          sig=sqrt(log(d43[0]/d32[0]));
          printf("%lf\n",1.0-0.57*exp(-sig)+0.2135*exp(-0.57*sig/0.2135)
           +0.0019*(cos(2.0*3.14159265*(1-exp(-0.75*pow(sig,0.7)
           -0.025*sig*sig*sig*sig)))-1.0));
          return(0);
        }
//-------------------------------------------------------------
      } else {
//      Read in all the data, which consists of triples (d43,d32,vol):
        if(verbose){
          printf("Please input data on the %i lognormal populations,",p);
          printf(" in the format of\n  d43 d32 volume \ntriples\n");
        }
//      We need to check that at least one volume is non-zero:
        zero_volume_flag=1;
        for(i=0;i<p;i++){
          scanf("%lf %lf %lf",&d43[i],&d32[i],&vol[i]);
          if(vol[i] > 0.0){zero_volume_flag=0;}
          if(d43[i] <= 0.0 || d32[i]<=0.0 || d32[i]>d43[i]){
            fprintf(stderr,"Error: need 0 < d32 < d43.\n");
            return(-1);
          }
          if(vol[i] < 0.0){
            fprintf(stderr,"Error: need positive values for the volume.\n");
            return(-1);
          }
        }
      }
    }
    if(lflag==2){
//    This is the case where the sphere size distributino is a sum
//    of lognormal distributions:
      if(p==1 && uflag==0){
//      Read in the data, which consists of the pairs sigma alone
        if(verbose){
          printf("Please input the log width sigma for the");
          printf(" lognormal population.\n",p);
        }
        scanf("%lf",&sigma[0]);
        vol[0]=1.0;
        d43[0]=1.0;
        if(sigma[0]<0.0){
          fprintf(stderr,"Error: must have sigma >= 0.\n");
          return(-1);
        }
//      Now deal with the -s case:
//------This piece of code uses a simple analytic approximation
//      for the case of one lognormal distribution:
        if(sflag){
          if(verbose){printf("\nPredicted maximum packing fraction:\n");}
          sig=sigma[0];
          printf("%lf\n",1.0-0.57*exp(-sig)+0.2135*exp(-0.57*sig/0.2135)
           +0.0019*(cos(2.0*3.14159265*(1-exp(-0.75*pow(sig,0.7)
           -0.025*sig*sig*sig*sig)))-1.0));
          return(0);
        }
//-------------------------------------------------------------
      } else {
//      Read in all the data, which consists of pairs (sigma,vol):
        if(verbose){
          printf("Please input data on the %i lognormal populations,",p);
          printf(" in the format of\n d43 sigma volume \ntriples\n");
        }
//      We need to check that at least one volume is non-zero:
        zero_volume_flag=1;
        for(i=0;i<p;i++){
          scanf("%lf %lf %lf",&d43[i],&sigma[i],&vol[i]);
          if(vol[i] > 0.0){zero_volume_flag=0;}
          if(sigma[i] < 0.0 || d43[i]<=0.0){
            fprintf(stderr,"Error: need sigma >=0 and d43 > 0.\n");
            return(-1);
          }
          if(vol[i] < 0.0){
            fprintf(stderr,"Error: need positive values for the volume.\n");
            return(-1);
          }
        }
      }
    }
    if(lflag==0){
//    This is the case where the sphere size distribution is a sum
//    of delta functions:
//
//    First deal with the -s case:
      if(sflag){printf("%lf\n",0.6435);return(0);}
//    Then all the other cases:
      if(p==1 && (1-uflag)){
        d43[0]=1.0;
        vol[0]=1.0;
      } else {
//      Read in all the data, which consists of pairs (d43,vol):
        if(verbose){
          printf("Please input data on the %i delta function populations,",p);
          printf(" in the format of\n  diameter volume \npairs\n");
        }
//      We need to check that at least one volume is non-zero:
        zero_volume_flag=1;
        for(i=0;i<p;i++){
          scanf("%lf %lf",&d43[i],&vol[i]);
          if(vol[i] > 0.0){zero_volume_flag=0;}
          if(d43[i] <= 0.0){
            fprintf(stderr,"Error: need strictly positive values for");
            fprintf(stderr," the diameter.\n");
            return(-1);
          }
          if(vol[i] < 0.0){
            fprintf(stderr,"Error: need positive values for the volume.\n");
            return(-1);
          }
//        Check that the sizes are distinct, and if not, then add the current
//        size to the earlier one which is the same, and reduce p by 1:
          for(j=0;j<i;j++){
            if(d43[j]==d43[i]){
              vol[j]=vol[j]+vol[i];
              p--;
              i--;
              break;
            }
          }
        }
        if(zero_volume_flag){
          fprintf(stderr,"Error: One of the volumes must be greater");
          fprintf(stderr," than zero.\n");
          return(-1);
        }
      }
    }

//  Now get the list of a's and the normalization variable 'norm':
    if(verbose){printf("Getting the a values:\n");}
    get_a_and_other_things();

//  Then construct the list of rod lengths:
    if(verbose){printf("Constructing the list of rod lengths:\n");}
    getL();


    if(verbose){printf("\nPredicted maximum packing fraction:\n");}
    printf("%lf\n",phimax());
}

\end{verbatim}
\end{tiny}

\section{Appendix: Code for the `KTS' algorithm}
This is an implementation of the algorithm from Ref. \cite{KTS},
which was used to generate data in figures \ref{lognormal_compare},
\ref{size_dist_images} and \ref{distributions_compare},
and table \ref{KTStable}.

Again, if you wish to use this code, it is most easily obtained 
from the \LaTeX\ file used to generate the document you are reading.
Save the code as a file {\tt "simple\_KTS.c"} then compile using,
for example {\tt "gcc simple\_KTS.c -o simple\_KTS -lm -O3"}.
The program will ask for the data it needs (including a list of
sphere sizes), but in practice this is best supplied by generating
the input in a file beforehand. Two files will be generated as output:
a general file, which contains data on the volume fraction as the
simulation progresses, and a `povray' file, which can be used
as input to the open source ratracing program POV-Ray (http://www.povray.org),
in order to display an image of the packing.

\begin{tiny}
\begin{verbatim}

/*-------------------------------------------------------------*
*                         simple_KTS                           *
*                                                              *
*  A program to estimate the maximum packing fraction of       *
*  hard spheers with various distributions of diameters,       *
*  based on a hard sphere algorithm "KTS" published in:        *
*  Journal of Chemical Physics, 117, 8212 (2002).              *
*  Packing happens in a box with periodic boundary conditions  *
*  and of size 1 in each direction. This code is a completely  *
*  naieve implementation of the published KTS algorithm, with  *
*  no optimizations. Most particularly, neighbour checking     *
*  is done in a simple, n^2 manner.                            *
*                                                              *
*  Author: Robert Farr.                                        * 
*                                                              * 
*  Released under the GNU General Public Licence, version 2.   *
*                                                              * 
*-------------------------------------------------------------*/

#include<stdio.h>
#include<math.h>

double phi_target;
double x[20000][3];
double v[20000][3];
double r0[20000];
// The actual radii at time t, are given by r0*(1+delta*t)
double delta=0.1;
double t;
// The next two spheres to collide:
int p1_next,p2_next;
int N;
int debug=0;
int end_flag=0;

void setup_radii(){
 // Set up the Number of spheres, their initial radii, and
 // re-size them so that they take up little enough volume to do RSA
 int i,flag;
 double vol,norm,pi;
 t=0.0;
 pi=3.14159265;
 vol=0.0;
 flag=1;
 while(flag){
  printf("How many spheres? ");
  scanf("%i",&N);
  flag=0;
  if(N<2 || N>=20000){
   printf("N must be in the range 2 to 19999\n");
   flag=1;
  }
 }
 printf("What is the volume fraction for the initial random\n");
 printf("placement? (suggested value 0.15, but smaller values\n");
 printf("may be necessary if the initial placement fails): ");
 scanf("%lf",&phi_target);
 for(i=0;i<N;i++){
  printf("What is the radius of sphere number %i? ",i);
  scanf("%lf",&r0[i]);
  vol+=(4.0*pi/3.0)*pow(r0[i],3);
 }
 for(i=0;i<N;i++){
  printf("%lf ",r0[i]);
 }
 printf("\n");
 // resize the spheres so that they have the target volume fraction:
 for(i=0;i<N;i++){
  r0[i]=r0[i]*pow(phi_target/vol,1.0/3.0);
 }
}

int displace(double a,double b){
 // How many units we need to displace b, in order to be
 // closest to a
 return(-(int)floor(b-a+0.5));
}

double min_dist(double a,double b){
 // The distance from a to b (i.e. b relative to a) when
 // the mapping to the nearest image is done
 return((b-a)-floor(b-a+0.5));
}

void RSA(){
 // Place sphere by random sequential absorption.
 // For best results, the sphere order should have been randomized.
 int i,ii,ic;
 double dist2;
 int flag;
 printf("RSA placement\n");
 for(i=0;i<N;i++){
  flag=1;
  printf("Place %i\n",i);
  while(flag){
   // Place a sphere:
   for(ic=0;ic<3;ic++){
    x[i][ic]=1.0e-6*(rand()%1000000);
   }
   flag=0;
   // Look for collisions:
   for(ii=0;ii<i;ii++){
    dist2=0.0;
    for(ic=0;ic<3;ic++){
     dist2+=min_dist(x[i][ic],x[ii][ic])*min_dist(x[i][ic],x[ii][ic]);
    }
    if(dist2<(r0[i]+r0[ii])*(r0[i]+r0[ii])*(1.0+delta*t)*(1.0+delta*t)){
     flag=1;
    }
   }
  }
 }
 printf("...done\n");
}

void re_image(){
 // Map all the spheres into a single periodic image of the system
 int i,j;
 for(i=0;i<N;i++){
  for(j=0;j<3;j++){
   while(x[i][j]<0.0){x[i][j]+=1.0;}
   while(x[i][j]>=1.0){x[i][j]-=1.0;}
  }
 }
}

void re_order(){
 int i,choose;
 double r0_store[20000];
 int re_order_flag[20000];
 printf("Initial re-ordering of the spheres randomly at start\n");
 // This just re-orders the initial radii, so can only be done
 // right at the start (after the radii are assigned).
 for(i=0;i<N;i++){
  r0_store[i]=r0[i];
  re_order_flag[i]=0;
 }
 for(i=0;i<N;i++){
  choose=rand()%N;
  while(re_order_flag[choose]){choose=rand()%N;}
  re_order_flag[choose]=1;
  r0[i]=r0_store[choose];
 }
 printf("...re-order finished.\n");
}

void print_radii(){
 int i;
 for(i=0;i<N;i++){
  printf("%lf\n",r0[i]);
 }
}

double time_to_collision(int p1,int p2){
 int ic;
 int flag;
 double v_rel[3];
 double x_rel[3];
 double xx,vv,xv,rr,r0r0;
 double dt;
 double work;
 // Terms in the quadratic equation:
 double a,b,c;
 for(ic=0;ic<3;ic++){
  v_rel[ic]=v[p2][ic]-v[p1][ic];
  // Find the closest image:
  x_rel[ic]=min_dist(x[p1][ic],x[p2][ic]);
 }
 vv=0.0;
 xx=0.0;
 xv=0.0;
 for(ic=0;ic<3;ic++){
  vv+=v_rel[ic]*v_rel[ic];
  xx+=x_rel[ic]*x_rel[ic];
  xv+=x_rel[ic]*v_rel[ic];
 }
 r0r0=(r0[p1]+r0[p2])*(r0[p1]+r0[p2]);
 rr=r0r0*(1.0+delta*t)*(1.0+delta*t);
 if(xx+1.0e-9<rr){
  printf("Jammed!!! %i %i r1r2 dist %lf %lf\n",p1,p2,sqrt(rr),sqrt(xx));
  end_flag=1;
 }
 // Now work out the coefficients in the quadratic equation for dt:
 a=vv-delta*delta*r0r0;
 b=2.0*xv-2.0*delta*r0r0*(1.0+delta*t);
 c=xx-r0r0*(1.0+delta*t)*(1.0+delta*t);
 // Now see if there is a collision, and if it happens at positive time:
 dt=1.0e+20;
 if(b*b-4.0*a*c>0.0){
  dt=(-b-sqrt(b*b-4.0*a*c))/(2.0*a);
  if(dt<0.0){dt=1.0e+20;}
 }
 return(dt);
}

void bounce(){
 double x_rel[3];
 double v_rel[3];
 // centre of mass velocity:
 double v_cm[3];
 // component of relative velocity parallel to the relative displacement
 double v_parallel;
 double v_perp[3];
 double dist;
 int p1,p2;
 int ic;
 if(debug)printf("bounce...\n");
 p1=p1_next;
 p2=p2_next;
 for(ic=0;ic<3;ic++){
  v_rel[ic]=v[p2][ic]-v[p1][ic];
  v_cm[ic]=0.5*(v[p2][ic]+v[p1][ic]);
  // Find the closest image:
  x_rel[ic]=min_dist(x[p1][ic],x[p2][ic]);
 }
 dist=sqrt(x_rel[0]*x_rel[0]+x_rel[1]*x_rel[1]+x_rel[2]*x_rel[2]);
 if(debug)printf(">>> dist, r1r2 %lf %lf\n",dist,(r0[p1]+r0[p2])*(1.0+delta*t));
 v_parallel=(x_rel[0]*v_rel[0]+x_rel[1]*v_rel[1]+x_rel[2]*v_rel[2])/dist;
 for(ic=0;ic<3;ic++){
  v_perp[ic]=v_rel[ic]-x_rel[ic]*v_parallel/dist;
 }
 // Print out the old velocities:
 if(debug)printf("v1: %lf %lf %lf\n",v[p1][0],v[p1][1],v[p1][2]);
 if(debug)printf("v2: %lf %lf %lf\n",v[p2][0],v[p2][1],v[p2][2]);
 // Assign new velocities (unchanged would be:
 //      v[p1][ic]=v_cm[ic]-0.5*(v_perp[ic]+x_rel[ic]*v_parallel/dist);
 //      v[p2][ic]=v_cm[ic]+0.5*(v_perp[ic]+x_rel[ic]*v_parallel/dist);
 for(ic=0;ic<3;ic++){
  v[p1][ic]=v_cm[ic]-0.5*(v_perp[ic]-x_rel[ic]*v_parallel/dist)
    -1.0*delta*(r0[p1]+r0[p2])*x_rel[ic]/dist;
  v[p2][ic]=v_cm[ic]+0.5*(v_perp[ic]-x_rel[ic]*v_parallel/dist)
    +1.0*delta*(r0[p1]+r0[p2])*x_rel[ic]/dist;
  //v[p1][ic]=0.0;
  //v[p2][ic]=0.0;
 }
 // Print out the new velocities:
 if(debug)printf("v1: %lf %lf %lf\n",v[p1][0],v[p1][1],v[p1][2]);
 if(debug)printf("v2: %lf %lf %lf\n",v[p2][0],v[p2][1],v[p2][2]);
 if(debug)printf("...bounce done.\n");
}

double min_time_to_collision(){
 int p1,p2;
 double dt,dtwork;
 if(debug)printf("min time to collision...\n");
 dt=1.0e+20;
 for(p1=0;p1<N;p1++){
  for(p2=0;p2<p1;p2++){
   dtwork=time_to_collision(p1,p2);
   if(dtwork<dt){
    dt=dtwork;
    p1_next=p1;
    p2_next=p2;
   }
  }
 }
 if(debug)printf("p1_next, p2_next %i %i\n",p1_next,p2_next);
 if(debug)printf("...done.\n");
 return(dt);
}

void povray_output(char povray_file[200]){
 FILE *f1;
 int i;
 f1=fopen(povray_file,"w");
 fprintf(f1,"\
#include \"colors.inc\"\n\
#include \"shapes.inc\"\n\
#include \"textures.inc\"\n\
#include \"finish.inc\"\n\
\n\
background{White}\n\
camera {  \n\
  location  <5,3,-10>\n\
  look_at  <0,0,0>\n\
  angle 10\n\
}\n\
\n\
#declare sph=union{\n");
 for(i=0;i<N;i++){
  fprintf(f1,"sphere{<%lf,%lf,%lf>,%lf}\n",x[i][0],x[i][1],
     x[i][2],r0[i]*(1.0+delta*t));
 }
 fprintf(f1,"\n}\n\
object{sph\n\
 texture{\n\
  pigment{color rgb <0.7,0.7,0.7>}\n\
  finish{ambient 0.2 diffuse 0.4 specular 0.6 roughness 0.05}\n\
 }\n\
 translate <-0.5,-0.5,-0.5>\n\
}\n\
\n\
light_source {<500,600,-1000> color White}\n\
");
 fclose(f1);
}

void positions_output(){
 FILE *f3;
 int i;
 f3=fopen("positions.dat","w");
 fprintf(f3,"%i\n",N);
 for(i=0;i<N;i++){
  fprintf(f3,"%lf %lf %lf %lf\n",x[i][0],x[i][1],x[i][2],r0[i]*(1.0+delta*t));
 }
 fclose(f3);
}


void setup_velocities(){
 int i,coord,flag;
 double norm;
 for(i=0;i<N;i++){
  flag=1;
  while(flag){
   norm=0.0;
   for(coord=0;coord<3;coord++){
    v[i][coord]=(2.0e-6*(rand()%1000000))-1.0;
    norm+=v[i][coord]*v[i][coord];
   }
   if(norm<1.0){
    flag=0;
   }
  }
 }
}

void renormalize_velocities(){
 int i,coord;
 double norm,mean_norm;
 mean_norm=0.0;
 for(i=0;i<N;i++){
  norm=0.0;
  for(coord=0;coord<3;coord++){
   norm+=v[i][coord]*v[i][coord];
  }
  mean_norm+=sqrt(norm);
 }
 mean_norm=mean_norm/N;
 if(debug)printf("mean speed before renormalization=%lf\n",mean_norm);
 for(i=0;i<N;i++){
  for(coord=0;coord<3;coord++){
   v[i][coord]=v[i][coord]/mean_norm;
  }
 }
}

void move_spheres(double dt){
 int i,ic;
 for(i=0;i<N;i++){
  for(ic=0;ic<3;ic++){
   x[i][ic]+=dt*v[i][ic];
  }
 }
}

void check_overlaps(){
 int p1,p2;
 int i[3];
 double dist;
 int ic;
 for(p1=0;p1<N;p1++){
  for(p2=0;p2<p1;p2++){
   for(i[0]=-1;i[0]<=1;i[0]++){
   for(i[1]=-1;i[1]<=1;i[1]++){
   for(i[2]=-1;i[2]<=1;i[2]++){
    dist=0.0;
    for(ic=0;ic<3;ic++){
     dist+=(x[p1][ic]-x[p2][ic]+i[ic])*(x[p1][ic]-x[p2][ic]+i[ic]);
    }
    dist=sqrt(dist);
    if(dist<(r0[p1]+r0[p2])*(1.0+delta*t)){printf("! %i %i\n",p1,p2);}
   }
   }
   }
  }
 }
}

double phi(){
 int i;
 double phi;
 double pi=3.14159265;
 phi=0.0;
 for(i=0;i<N;i++){
  phi+=(4.0*pi/3.0)*pow(r0[i]*(1.0+delta*t),3);
 }
 return(phi);
}

void main(){
 int it;
 int iter,iter_max;
 int iseed;
 double dt,dt_av;
 FILE *f2;
 char output_file[200];
 char povray_file[200];
 iseed=(unsigned int)time(NULL);
 srand(iseed);
 printf("File name for general output? ");
 scanf("%s",output_file);
 printf("File name for povray output? ");
 scanf("%s",povray_file);
 f2=fopen(output_file,"w");
 setup_radii();
 re_order();
 RSA();
 povray_output(povray_file);
 check_overlaps();
 setup_velocities();
 renormalize_velocities();
 // Now go through the cycle:
 it=0;
 iter_max=1000;
 fprintf(f2,"# N=%i\n",N);
 fprintf(f2,"# iseed=%i\n",iseed);
 fprintf(f2,"# phi_target=%lf\n",phi_target);
 fprintf(f2,"#\n# Iteraton number, phi, average of dt, t\n");
 fflush(f2);
 while(!end_flag){
  dt_av=0.0;
  for(iter=0;iter<iter_max;iter++){
   dt=min_time_to_collision();
   if(debug)printf("dt=%g\n",dt);
   t+=dt;
   dt_av=dt_av+dt/iter_max;
   it++;
   move_spheres(dt);
   if(debug){
    printf("check overlap after moving:\n");
    check_overlaps();
   }
   bounce();
   re_image();
   renormalize_velocities();
  }
  printf("%g %g %g %g\n",1.0*it,phi(),dt_av,t);
  fprintf(f2,"%g %g %g %g\n",1.0*it,phi(),dt_av,t);
  fflush(f2);
  povray_output(povray_file);
  positions_output();
 }
 povray_output(povray_file);
 fclose(f2);
}
\end{verbatim}
\end{tiny}


\begin{thebibliography}{1}
\bibitem{KD} I. M. Krieger, {\em Adv. Col. Int. Sci.} {\bf 3}, 111 (1972).
\bibitem{Barnes} H. A. Barnes, J. F. Hutton and K. Walters, `An Introduction to Rheology', Elsevier, 1989.
\bibitem{Ackerson} Ackerson, B. J., J. Rheol. {\bf 34}, 553 (1990).
\bibitem{Mason} T. G. Mason, J. Bibette and D. A. Weitz, {\em J. Col. Int. Sci.} {\bf 179}, 439 (1996).
\bibitem{Foudazi} R. Foudazi, I. Masalova, and A. Ya. Malkin, J. Rheol. {\rm 56}, 1299 (2012).
\bibitem{Bernal} J. D. Bernal, J. Mason, {\em Nature} {\bf 188}, 910 (1960).
\bibitem{Torquato} S. Torquato, T. M. Truskett, and P. G. Debenedetti, Phys. Rev. Lett. {\bf 84}, 2064 (2000).
\bibitem{Kepler} J. Kepler, `Strena seu de Nive sexangula' (essay, 1611). Available in English translation as `The six-cornered snowflake', C. Hardie, Oxford Clarendon Press (1966) and online at http://www.thelatinlibrary.com/kepler/strena.html
\bibitem{Hales} T. C. Hales and S. P. Ferguson, Discrete \& Computational Geometry, {\bf 36}(1), 21 (2006).
\bibitem{Silbert} C. S. O’Hern, L. E. Silbert, A. J. Liu and Sidney R. Nagel, Phys. Rev. E {\bf 68}, 011306 (2003).
\bibitem{LuSt} B. D. Lubachevsky and F. H. Stillinger, J. Stat. Phys. {\bf 60}, 561 (1990).
\bibitem{Warren} P. Espanol and P. B. Warren, Europhysics Letters, {\bf 30}(4),191 (1995).
\bibitem{FG} R. S. Farr and R. D. Groot, J. Chem. Phys. {\bf 131} 244104 (2009).
\bibitem{Song} C. Song, P. Wang, and H. A. Makse, Nature London 453, 629 (2008).
\bibitem{Delaney} G. W. Delaney and P. W. Cleary, EPL {\bf 89} 34002 (2010).
\bibitem{Ouchiyama} N. Ouchiyama, T. Tanaka, {\em Ind. Eng. Chem. Fundam.} {\bf 20}, 66 (1981).
\bibitem{Biazzo} I. Biazzo, F.  Caltagirone, G. Parisi, F. Zamponi, {\em Phys. Rev. Lett.} {\bf 102}, 195701 (2009).
\bibitem{Danisch} M. Danisch, Y. Jin and H. A. Makse, Phys. Rev. E {\bf 81}, 051303 (2010).
\bibitem{KTS} A. R. Kansal, S. Torquato, and F. H. Stillinger, J. Chem. Phys. {\bf 117}, 8212 (2002).
\bibitem{Rajagopal} E. S. Rajagopal, {\em Kolloid Zh.}, {\bf 162}, 85 (1959).
\bibitem{Hollingsworth} K. G. Hollingsworth and M. I. Johns, {\em J. Col. Int. Sci.} {\bf 258}, 383 (2003).
\bibitem{Rosin} P. Rosin, {\em J. Inst. Fuel} {\bf 7}, 29 (1933).
\bibitem{Kondolf} G. M. Kondolf and A. Adhikari, {\em J. Sedimentary Res.}, {\bf 70}(3), 456 (2000).
\bibitem{Cormen} T. H. Cormen {\em et al.}, Introduction to Algorithms (MIT Press, 2009).
\bibitem{Arch} Marcus Vitruvius Pollio, {\em De architectura} (British Library Manuscript `Harley 2767'). Also available in translation by M. H. Morgan, {\em `Vitruvius: the ten books on architecture'}, Book IX (Harvard University Press, 1914), and at http://www.gutenberg.org/files/20239/20239-h/29239-h.htm
\bibitem{C} B. Kernighan and D. Ritchie, The C Programming Language (Prentice Hall, Inc., 1988).
\bibitem{GPL} http://www.gnu.org/licenses/old-licenses/gpl-2.0.html
\bibitem{SF} http://sourceforge.net/projects/spherepack1d/
\end{thebibliography}
\end{document}